\documentclass[a4paper,10pt,showpacs,showkeys,notitlepage]{revtex4-1}
\usepackage{amsmath,amssymb,textcomp,float,dsfont}
\usepackage[utf8]{inputenc}
\usepackage{amsmath}
\usepackage{amsfonts}
\usepackage{amssymb}
\usepackage{times,fullpage}
\usepackage{comment,graphicx}
\usepackage{array}

\graphicspath{{./figures/}}

\begin{document}

\newcommand{\nwc}{\newcommand}
\nwc{\vs}{\vspace}
\nwc{\hs}{\hspace}
\nwc{\la}{\langle}
\nwc{\ra}{\rangle}
\nwc{\nn}{\nonumber}
\nwc{\Ra}{\Rightarrow}
\nwc{\wt}{\widetilde}
\nwc{\lw}{\linewidth}
\nwc{\ft}{\frametitle}
\nwc{\ben}{\begin{enumerate}}
\nwc{\een}{\end{enumerate}}
\nwc{\bit}{\begin{itemize}}
\nwc{\eit}{\end{itemize}}
\nwc{\dg}{\dagger}
\nwc{\mA}{\mathcal A}
\nwc{\mD}{\mathcal D}
\nwc{\mB}{\mathcal B}

\nwc{\Tr}[1]{\underset{#1}{\mbox{Tr}}~}
\nwc{\pd}[2]{\frac{\partial #1}{\partial #2}}
\nwc{\ppd}[2]{\frac{\partial^2 #1}{\partial #2^2}}
\nwc{\fd}[2]{\frac{\delta #1}{\delta #2}}
\nwc{\pr}[2]{K(i_{#1},\alpha_{#1}|i_{#2},\alpha_{#2})}
\nwc{\av}[1]{\left< #1\right>}

\nwc{\zprl}[3]{Phys. Rev. Lett. ~{\bf #1},~#2~(#3)}
\nwc{\zpre}[3]{Phys. Rev. E ~{\bf #1},~#2~(#3)}
\nwc{\zpra}[3]{Phys. Rev. A ~{\bf #1},~#2~(#3)}
\nwc{\zjsm}[3]{J. Stat. Mech. ~{\bf #1},~#2~(#3)}
\nwc{\zepjb}[3]{Eur. Phys. J. B ~{\bf #1},~#2~(#3)}
\nwc{\zrmp}[3]{Rev. Mod. Phys. ~{\bf #1},~#2~(#3)}
\nwc{\zepl}[3]{Europhys. Lett. ~{\bf #1},~#2~(#3)}
\nwc{\zjsp}[3]{J. Stat. Phys. ~{\bf #1},~#2~(#3)}
\nwc{\zptps}[3]{Prog. Theor. Phys. Suppl. ~{\bf #1},~#2~(#3)}
\nwc{\zpt}[3]{Physics Today ~{\bf #1},~#2~(#3)}
\nwc{\zap}[3]{Adv. Phys. ~{\bf #1},~#2~(#3)}
\nwc{\zjpcm}[3]{J. Phys. Condens. Matter ~{\bf #1},~#2~(#3)}
\nwc{\zjpa}[3]{J. Phys. A ~{\bf #1},~#2~(#3)}
\nwc{\zpjp}[3]{Pramana J. Phys. ~{\bf #1},~#2~(#3)}

\title{ Interacting Multi-particle Classical Szilard Engine }
\author{P. S. Pal$^{a,b}$\email{} and A. M. Jayannavar$^{a,b}$\email{}} 
\email{priyo@iopb.res.in, jayan@iopb.res.in}
\affiliation{$^a$Institute of Physics, Sachivalaya Marg, Bhubaneswar-751005, India\\$^b$Homi Bhabha National Institute, Training School Complex, Anushakti Nagar, Mumbai 400085, India}
\begin{abstract}
Szilard engine(SZE) is one of the best example of how information can be used to extract work from a system. Initially, the working substance of SZE was considered to be a single particle. Later on, 
researchers has extended the studies of SZE to multi-particle systems and even to quantum regime. Here we present a detailed study of classical SZE consisting $N$ particles with inter-
particle interactions, i.e., the working substance is a low density non-ideal gas and compare the work extraction with respect to SZE with non-interacting multi particle system as working substance. 
We have considered two cases of interactions namely: (i) hard core interactions and (ii) square well interaction. Our study reveals that work extraction is less when more particles are interacting 
through hard core interactions. More work is extracted when the particles are interacting via square well interaction. Another important result for the second case  is that as we increase the particle
number the work extraction becomes independent of the initial position of the partition, as opposed to the first case. Work extraction depends crucially on the initial position of the partition. More 
work can be  extracted with larger number of particles when partition is inserted at positions near the boundary walls. 
\end{abstract}


\maketitle
\section{Introduction}
In 1929, Leo Szilard put forward a classic demonstration of Maxwell's demon \cite{max,vedral_09} by describing an engine that uses the information acquired by measurement of the state of the working substance to perform
some work \cite{Szilard}. Szilard considered a single particle in a box and executed the following four steps to extract work from the system:(a) insert a partition at the middle of the box, (b) measure the whether
the particle is in the right or left half, (c) attach the box with a bath at inverse temperature $\beta$ and allow the partition to move isothermally and quasistatically till the end of the box and (d) remove the 
partition so that the system regains its initial state. Now here arises a crucial question regarding the compatibility between this cyclic thermodynamic process and the second law of thermodynamics.
Apparently, at first stance one might think that work has been extracted out of nowhere. But it is not the case. It is now a widely known fact that the measurement process including erasure has an 
entropic cost which saves the second law \cite{landauer_61,landauer_91}. Employment of information to extract work out of a system has led to some insightful physics in the area of non-equilibrium
thermodynamics \cite{sagawa_08,mandal_12,parrondo_15,rana_16,rana_1611}. 

Although Szilard engine has been initially pictured in classical framework with single particle, recently it has been translated to quantum regime along with multi-particle scenario  and asymmetric 
effects \cite{kim_11,Lu_12,kim_16,pal_16}. In this work we considered a multi-particle classical Szilard engine, but these particle does not behave like an ideal gas. Instead we take $N$ particle non-ideal gas as our working substance
i.e., the particles have interactions between them. Here we have assumed the gas to be of low density so that we can safely use the equation of state of the gas upto second virial co-efficient. We 
worked on two types of interactions: (i) hard core interaction and (ii) square well interaction. Our study reveals that work extraction is less when more particles are interacting 
through hard core interactions. More work is extracted when the particles are interacting via square well interaction. Another important result for the second case  is that as we increase the particle
number the work extraction becomes independent of the initial position of the partition, as opposed to the first case. Work extraction depends crucially on the initial position of the partition.  More 
work can be  extracted with larger number of particles when partition is inserted at positions near the boundary walls. 

The gas is filled in a box of length $L$ and surface area $A$(total volume $V=A.L$). The partition
is inserted vetically at $l=xL$ ($0<x<1$).  Next the number of particle  is measured on the left half, say $n$. The system undergoes an isothermal expansion in contact with the heat bath and the final
 position of the partition be $l^n_{eq}$. Lastly the partition is removed. 
The work done by the SZE is given by \cite{kim_11,kim_16}
  \begin{equation}
  W_{tot}=-\frac{1}{\beta}\sum_{n=0}^N p_n\ln\left(\frac{p_n}{f_n}\right).
  \label{w_tot}
 \end{equation}
$\beta=(k_BT)^{-1}$, where $k_B$ denotes the Boltzmann constant which is taken to be unity in subsequent expressions. Here $p_n$ and $f_n$ are given by
 \begin{equation}
 p_n=\frac{Z_{n,N-n}(xL)}{\sum_{n'}Z_{n',N-n'}(xL)};f_n=\frac{Z_{n,N-n}(l^n_{eq})}{\sum_{n'}Z_{n',N-n'}(l^n_{eq})},
 \label{pf}
\end{equation}
 where $Z_{n,N-n}(X)=Z_n(X)Z_{N-n}(L-X)$ is a partition function that describes the situation of $n$ particles to the left of the partition and the remaining $N-n$ to  the right, in a 
thermal equilibrium. Physically, $p_n$ denotes the probability that there are $n$ particles to the left after partition and $f_n$ represents the probability to choose the case of $n$ particles on the 
left side of the wall when the wall is inserted at $l^n_{eq}$ in the time backward process. Note that one can choose $l$ freely when the wall is inserted while $l^n_{eq}$ is determined from the force 
(pressure) balance on both sides of the wall $F^L+F^R=0$. The whole problem now boils down to the calculation of the partition function for non-ideal gas in a confined volume for two interaction potentials,
as mentioned above. Section \ref{partition} deals with the partition function for non-ideal gases. Sections \ref{work_hc} and \ref{work_sw}  describes work extraction using non-ideal gases with 
hard-core potential and square well interactions respectively. A short discussion is added in Section \ref{dis}. Finally we conclude in Section \ref{conclusion}. Each section is made self consistent.

\section{Partition function of non-ideal gas}
\label{partition}
Partition function of non-ideal gas consisting of $N$ interacting particles confined  in a box of length $L$ and volume  $V$ at inverse temperature $\beta$ is given by 
\begin{equation}
 Z=Z_{id}\left[1+\frac{N^2}{2V}\int \{e^{-\beta U(\vec r)}-1\}d^3\vec r\right].
\end{equation}
Here $Z_{id}$ is the partition function for ideal gas and is  given by
\begin{equation}
 Z_{id}=\frac{cL^N}{N!},
\end{equation}
where $c$ is some constant.
$U(\vec r)$ is the interaction potential between two particles. If $U(\vec r)$ is taken to be spherically symmetric one can integrate out the angular degrees of freedom resulting
\begin{eqnarray}
 Z&=&Z_{id}\left[1+\frac{N^2}{2V}4\pi\int r^2\{e^{-\beta U(r)}-1\}dr\right],\nn\\
 &=& \frac{cL^N}{N!}\left[1-\frac{N^2}{V}B_2(\beta)\right],
\end{eqnarray}
where $B_2(\beta)=-2\pi\int r^2\{e^{-\beta U(r)}-1\}dr$ is the second virial co-efficient that arises in the equation of state of the non-ideal gas. 

\subsection{Hard Core interaction}
In this model the interaction potential is given by
\begin{eqnarray}
 U(r)&=&\infty\hspace{2cm}r<\sigma,\nn\\
 &=&0\hspace{2cm}\sigma<r,
\end{eqnarray}
where $\sigma$ is the hard core radius. 
\begin{figure}[H]
 \begin{center}
  \includegraphics[height=2.5 in,width=2.5 in]{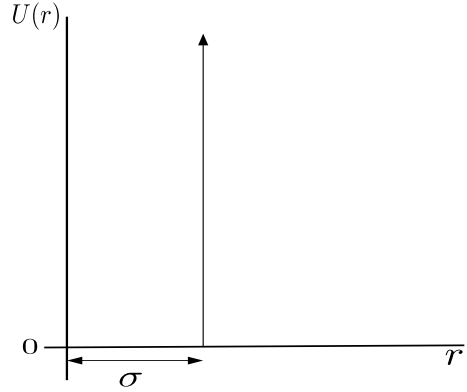}
  \caption{ Hard core potential.}
  \label{hard_core}
 \end{center}
\end{figure}
The expression for $B_2(\beta)$ is
\begin{equation}
 B_2(\beta)=2\pi\int_0^{\sigma} r^2dr=\frac{2}{3} \pi\sigma^3,
\end{equation}
and the correponding partition function is given by
\begin{equation}
 Z=\frac{cL^N}{N!}\left[1-\frac{2}{3V}\pi\sigma^3 N^2\right].
 \label{hc_z}
\end{equation}
Equation of state for such a non-ideal gas with hard core potential is 
\begin{equation}
 \beta P=\frac{N}{V-\frac{2}{3V}\pi\sigma^3 N},
 \label{eos_hc}
\end{equation}
where $P$ is the pressure of the gas. 
\subsection{ Square well interaction}
In this case the interaction potential is given by
\begin{eqnarray}
 U(r)&=&\infty\hspace{2 cm} r<\sigma\nn\\
 &=&-\varepsilon\hspace{2 cm}\sigma<r<R\sigma\nn\\
 &=&0\hspace{2 cm}R\sigma<r.
\end{eqnarray}
\begin{figure}[H]
 \begin{center}
  \includegraphics[height=2.5 in,width=2.5 in]{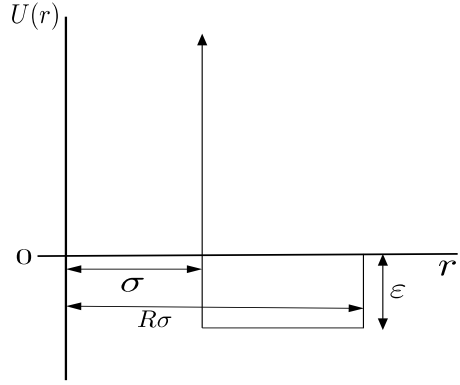}
  \caption{ Square well potential.}
  \label{square_well}
 \end{center}
\end{figure}
The expression for $B_2(\beta)$ is
\begin{eqnarray}
 B_2(\beta)&=&-2\pi\{-\int_0^{\sigma}r^2dr+(e^{\beta\varepsilon}-1)\int_{\sigma}^{R\sigma}r^2dr\}\nn\\
 &=&\frac{2}{3}\pi\sigma^3\{1-(R^3-1)(e^{\beta\varepsilon}-1)\}\nn\\
 &=&C_1-C_2e^{\beta\varepsilon},
\end{eqnarray}
where $C_1=\frac{2}{3}\pi\sigma^3R^3$ and $C_2=\frac{2}{3}\pi\sigma^3(R^3-1)$. The partition function is given by
\begin{equation}
Z=\frac{cL^N}{N!}\left\{1-\frac{N^2}{V}(C_1-C_2e^{\beta\varepsilon})\right\}. 
\label{sw_z}
\end{equation}
Corresponding equation of state for the non-ideal gas with square well interaction between particles is
\begin{equation}
 \beta P=\frac{N}{V}+B_2(\beta)\left(\frac{N}{V}\right)^2=\frac{N}{V}+(C_1-C_2e^{\beta\varepsilon})\left(\frac{N}{V}\right)^2.
 \label{pressure_sw}
\end{equation}
\section{Work extraction}
In classical systems, insertion and removal of partition does not involve work done on the system. Work extraction is only possible during the isothermal process and it is given by Eq.\ref{w_tot}.
The partition is initially inserted at $l=xL$ and during the isothermal process it shifts to $l=l^n_{eq}$ depending on the number of particles on the left/ right side. Since the partition 
function can now be calculated we can easily compute $p_n$, $f_n$  and the work for the above mentioned interactions.
\subsection{Hard core interaction}
\label{work_hc}
In this type of interaction between particles, $p_n$ is calculated using Eq.\ref{pf} and \ref{hc_z}
\begin{eqnarray}
 p_n&=&\frac{Z_{n,N-n}(xL)}{\sum_{n'}Z_{n',N-n'}(xL)}\nn\\
 &=&\frac{Z_{n}(xL)Z_{N-n}((1-x)L)}{\sum_{n'}Z_{n'}(xL)Z_{N-n'}((1-x)L)}\nn\\
 &=&\frac{{N\choose n}x^n(1-x)^{N-n}(1-\frac{\alpha n^2}{x})[1-\frac{\alpha (N-n)^2}{1-x}]}{\sum_{n'}{N\choose n'}x^{n'}(1-x)^{N-n'}(1-\frac{\alpha n'^2}{x})[1-\frac{\alpha (N-n')^2}{1-x}]}
\end{eqnarray}
Here $\alpha=\frac{2}{3V}\pi\sigma^3$.
Before the computation of $f_n$, we need to calculate the equilibrium length $l^n_{eq}$ upto which the partition moves due to the pressure difference  on two sides of the partition. 
For this we require pressure exerted on the partition by the gas on both sides to be same, i.e.,  $P^L=P^R$, which can be rewritten using the equation of state (Eq.\ref{eos_hc})as
\begin{eqnarray}
 \frac{n}{Al^n_{eq}-\frac{2}{3V}\pi\sigma^3 n}&=&\frac{N-n}{A(L-l^n_{eq})-\frac{2}{3V}\pi\sigma^3 (N-n)},\nn\\
 An(L-l^n_{eq})-\frac{2}{3V}\pi\sigma^3 (N-n)n&=&(N-n)Al^n_{eq}-\frac{2}{3V}\pi\sigma^3 n(N-n),\nn\\
 l^n_{eq}&=&\frac{n}{N}L.
\end{eqnarray}
 Now the expression of $f_n$ can be easily calculated and is given by
 \begin{equation}
  f_n=\frac{{N\choose n}(\frac{n}{N})^n(1-\frac{n}{N})^{N-n}(1-\alpha nN)[1-\alpha (N-n)N]}{\sum_{n'}{N\choose n'}(\frac{n}{N})^{n'}(1-\frac{n}{N})^{N-n'}(1-\alpha n'N)[1-\alpha (N-n')N]}
 \end{equation}
\begin{figure}[H]
 \begin{center}
  \includegraphics[height=2.5 in,width=2.5 in]{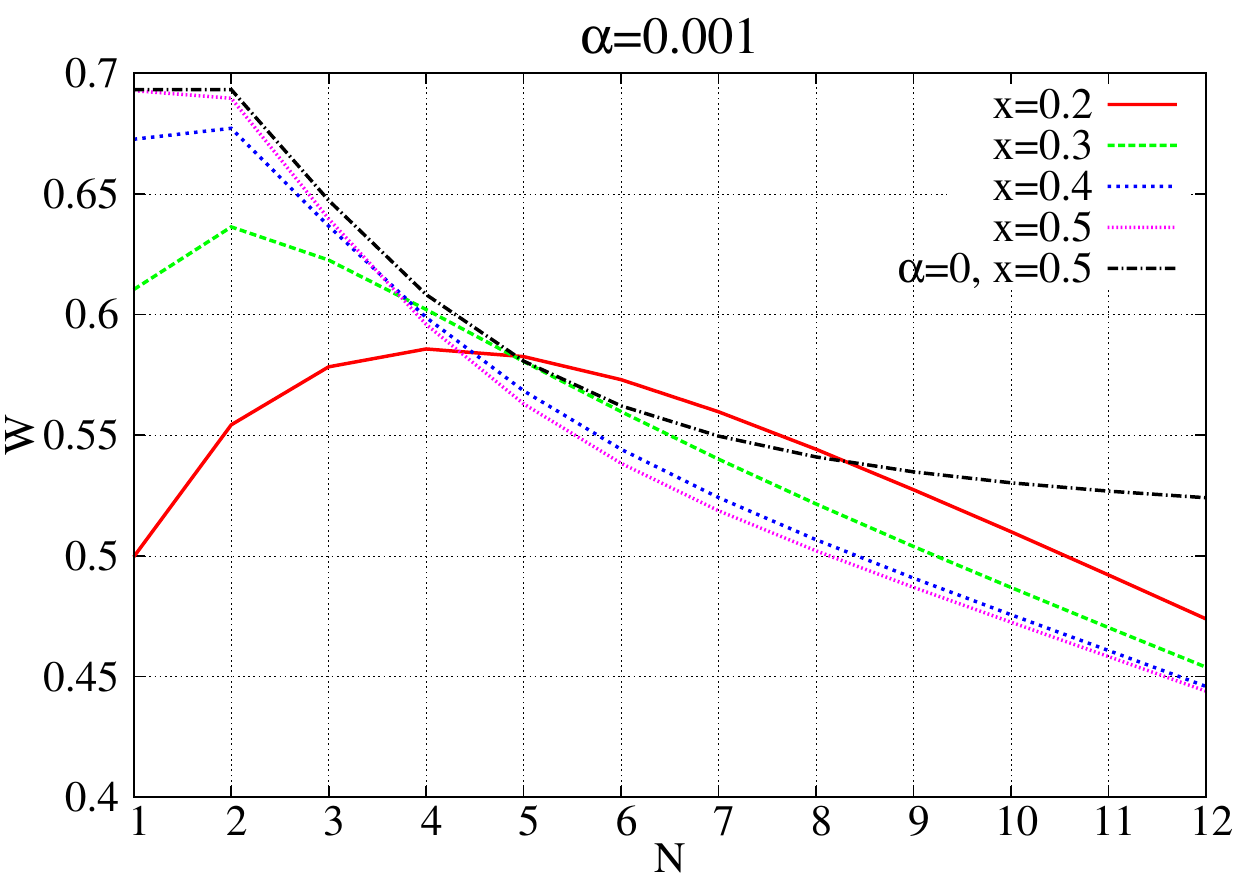}
  \caption{ Average work extraction as function of number of particles.}
  \label{W_vs_N}
 \end{center}
\end{figure}
Fig.\ref{W_vs_N} represents the average work as the function of number of particles with hard core interactions for different initial position of the partition.  Similar to the non-interacting case,
work extraction decreases when number of particles increases. For comparison, we have also included in the figure, the work extraction plot with non-interacting particles ($\alpha=0$) when the partition
is initially placed at the mid-point $x=0.5$.  As can be seen, work extraction lower when we are dealing with non-ideal gases rather than ideal gases.
\begin{figure}[H]
 \begin{center}
  \includegraphics[height=2.5 in,width=2.5 in]{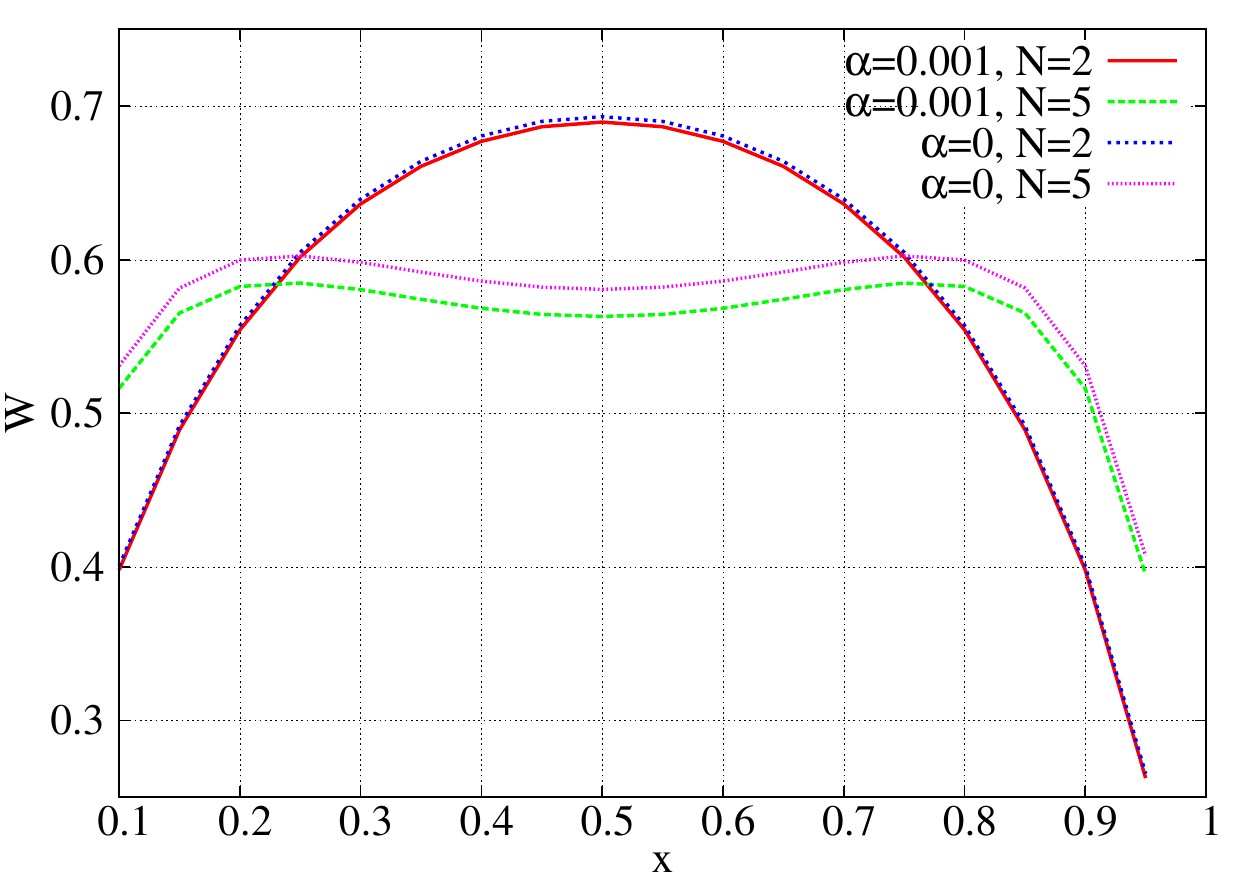}
  \caption{ Average work extraction as function of position of the partition.}
  \label{W_vs_x}
 \end{center}
\end{figure}
Fig.\ref{W_vs_x} depicts the behavior of average work as a function of the position where the partition is initially inserted. The plot clearly shows that interactions plays a major role in work 
extraction as the number of particles is increased. Inter-particle interactions decreases work extraction. Average work is symmetric about $x=0.5$ for obvious reason. Another point to be noted is that
work extraction is more for $x\neq 0.5$, as particle number is increased. 
\subsection{ Square well potential}
\label{work_sw}
\begin{figure}[H]
 \begin{center}
  \includegraphics[height=2.5 in,width=2.5 in]{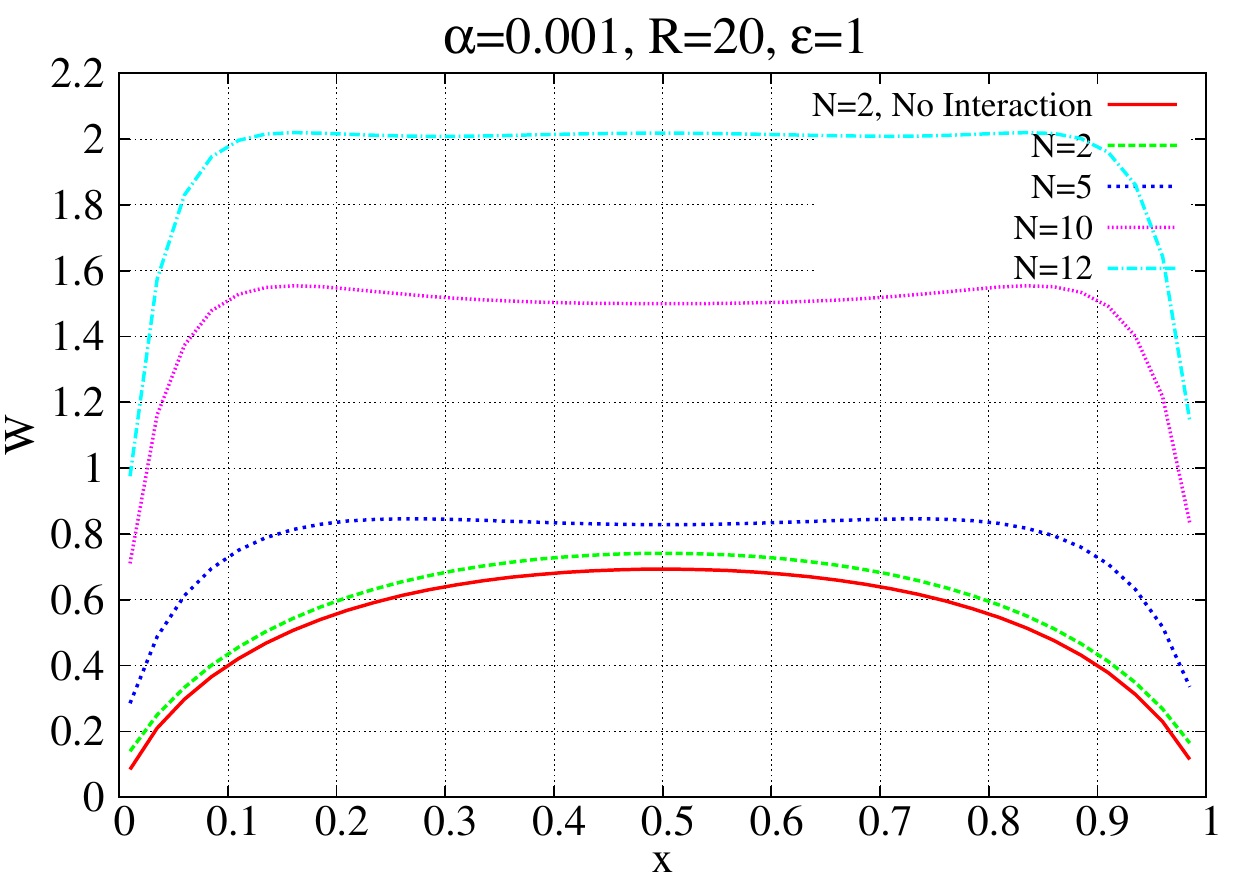}
  \caption{ Average work extraction as function of position of the partition.}
  \label{W_vs_x_sw}
 \end{center}
\end{figure}

In this type of interaction between particles, $p_n$ is calculated using Eq.\ref{pf} and \ref{sw_z}
\begin{eqnarray}
 p_n&=&\frac{Z_{n,N-n}(xL)}{\sum_{n'}Z_{n',N-n'}(xL)},\nn\\
 &=&\frac{Z_{n}(xL)Z_{N-n}((1-x)L)}{\sum_{n'}Z_{n'}(xL)Z_{N-n'}((1-x)L)},\nn\\
 &=&\frac{{N\choose n}x^n(1-x)^{N-n}(1-\frac{\gamma n^2}{x})[1-\frac{\gamma (N-n)^2}{1-x}]}{\sum_{n'}{N\choose n'}x^{n'}(1-x)^{N-n'}(1-\frac{\gamma n'^2}{x})[1-\frac{\gamma (N-n')^2}{1-x}]},
\end{eqnarray}
where $\gamma=C'_1-C'_2e^{\beta\varepsilon}$ with $C'_1=\frac{1}{V}C_1=\frac{2\pi\sigma^3R^3}{3V}=\alpha R^3$ and $C'_2=\frac{1}{V}C_2=\frac{2\pi\sigma^3(R^3-1)}{3V}=\alpha (R^3-1)$.
To calculate the equilibrium position of the partition  we need to equate the pressure (given by Eq.\ref{pressure_sw}) on the both sides of the partition and hence we need to solve the following 
equation:
\begin{equation}
 \frac{n}{Al^n_{eq}}+(C_1-C_2e^{\beta\varepsilon})\left(\frac{n}{Al^n_{eq}}\right)^2= \frac{N-n}{A(L-l^n_{eq})}+(C_1-C_2e^{\beta\varepsilon})\left[\frac{N-n}{A(L-l^n_{eq})}\right]^2.
\end{equation}
The physical solution of the above equation is $l^n_{eq}=(n/N)L$. 
%
Using Eq.\ref{pf}, $f_n$ can be calculated 
\begin{eqnarray}
 f_n&=&\frac{Z_{n,N-n}(l^n_{eq})}{\sum_{n'}Z_{n',N-n'}(l^n_{eq})},\nn\\
 &=&\frac{Z_{n}(l^n_{eq})Z_{N-n}(L-l^n_{eq})}{\sum_{n'}Z_{n'}(l^n_{eq})Z_{N-n'}(L-l^n_{eq})},\nn\\
 &=& \frac{{N\choose n}(\frac{n}{N})^n(1-\frac{n}{N})^{N-n}(1-\gamma nN)[1-\gamma (N-n)N]}{\sum_{n'}{N\choose n'}(\frac{n}{N})^{n'}(1-\frac{n}{N})^{N-n'}(1-\gamma n'N)[1-\gamma (N-n')N]}
\end{eqnarray} 
\begin{figure}[H]
 \begin{center}
  \includegraphics[height=2.5 in,width=2.5 in]{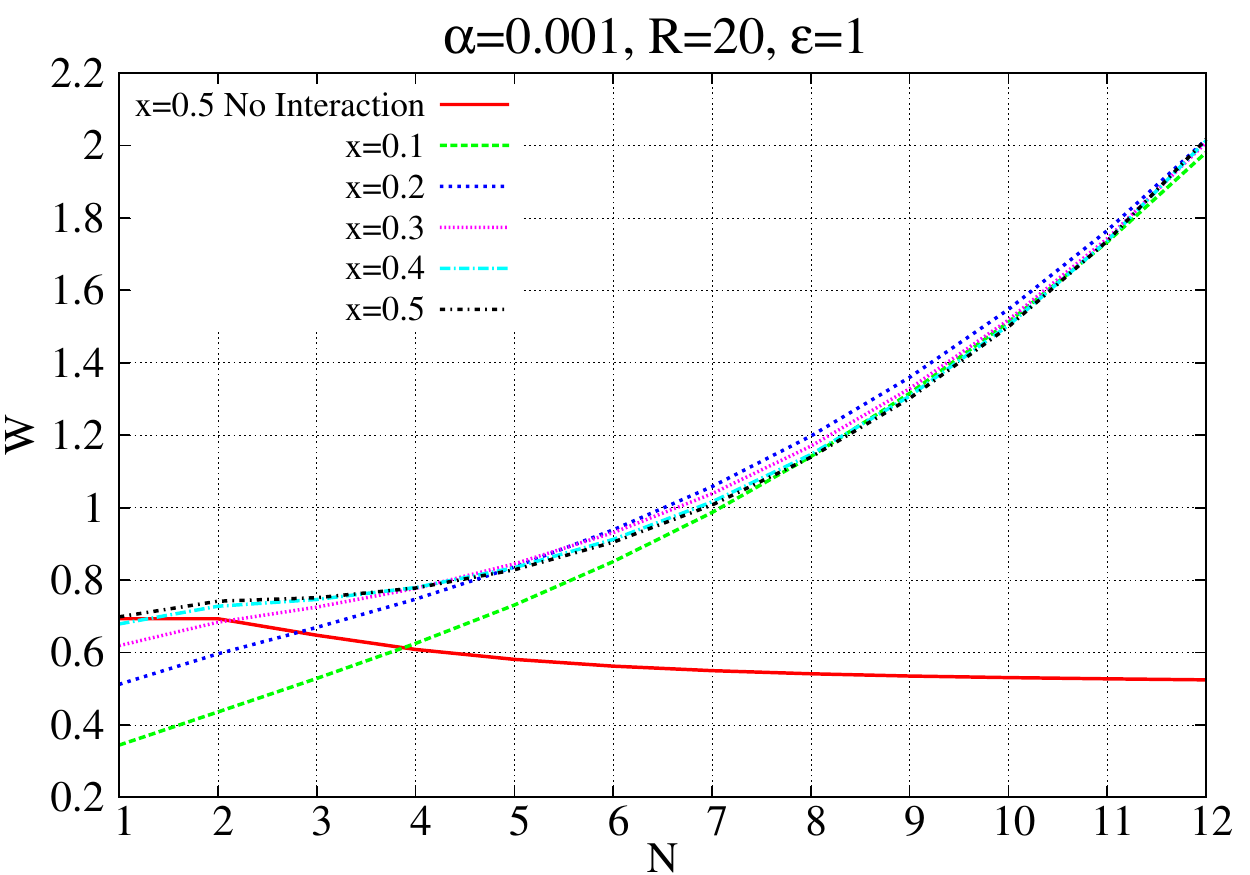}
  \caption{ Average work extraction as function of number of particles.}
  \label{W_vs_N_sw}
 \end{center}
\end{figure}
Fig.\ref{W_vs_N_sw} shows the behavior of average work extraction as we increase the number of particle. The plot shows that for small number of particles work extraction is less than the non-interacting
case. But with increasing particle number the more work is extracted using interacting particle system. Another important thing to notice is that work extraction is independent of the initial position
of the partition for higher values of $N$. This fact is supported by the plot given in Fig.\ref{W_vs_x_sw}, where the amount of work is plotted against $x$ - initial position of wall- for different
$N$ values. This graph depicts the fact that as one increase the number of particles, the maximum work extraction, which was initially at $x=0.5$, shifts to either sides and finally for $N=12$, the 
effect of asymmetry on work extraction becomes very less in the range $0.1\leq x\leq 0.9$.
\begin{figure}[H]
 \begin{center}
  \includegraphics[height=2.5 in,width=2.5 in]{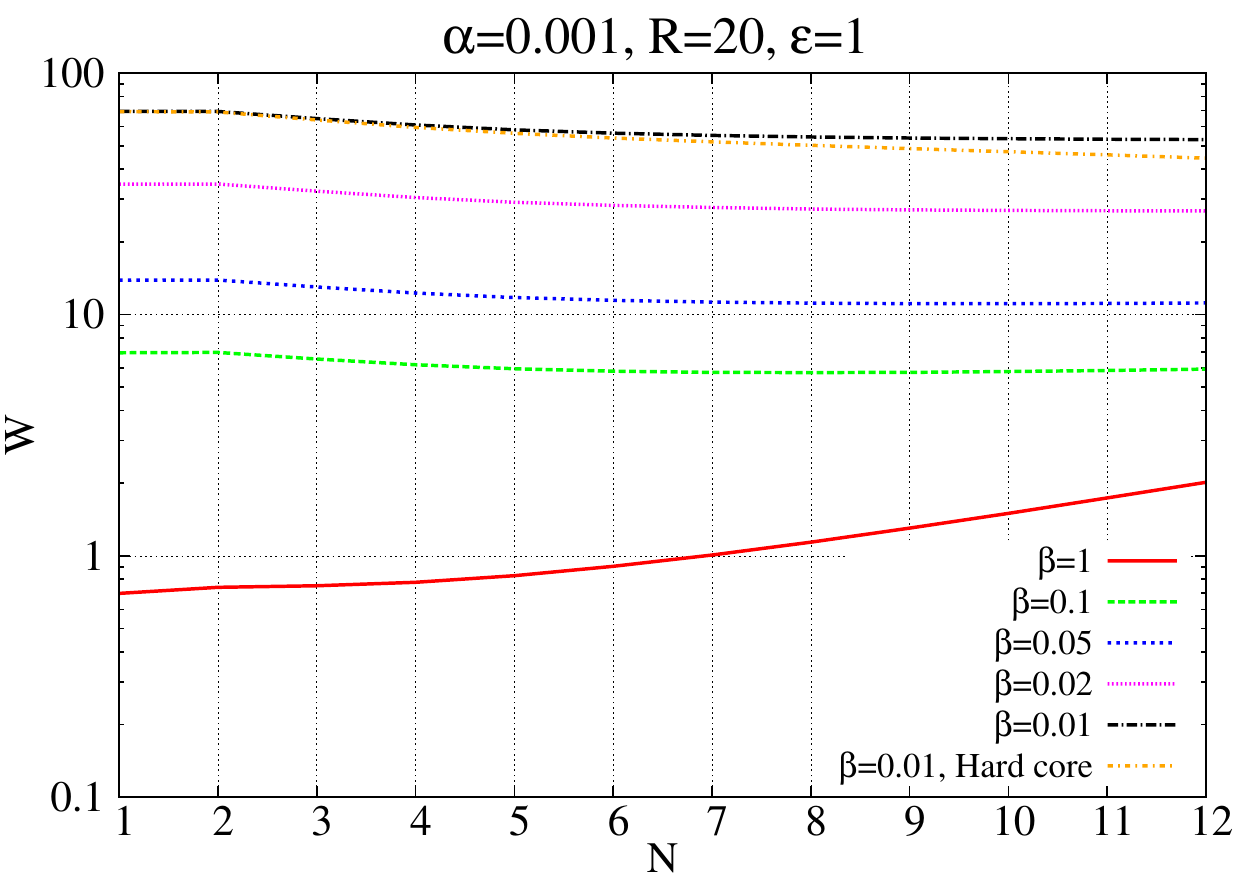}
  \caption{ Average work extraction as function of number of particles for different temperature values.}
  \label{W_vs_N_sw_T}
 \end{center}
\end{figure}
Fig.\ref{W_vs_N_sw_T} shows the behavior of average work at different temperatures in presence of square well interaction between particles. Work extraction increases as the temperature is increased.
At high enough temperatures the particles do not feel the negative well of the interaction and the working substance is expected to act as a non-ideal gas with hard core interaction. In the plot a 
comparison of work extraction  is shown between particles with square well interactions and particles with hard core interaction for $\beta=0.01$.
\begin{figure}[H]
 \begin{center}
  \includegraphics[height=2.5 in,width=2.5 in]{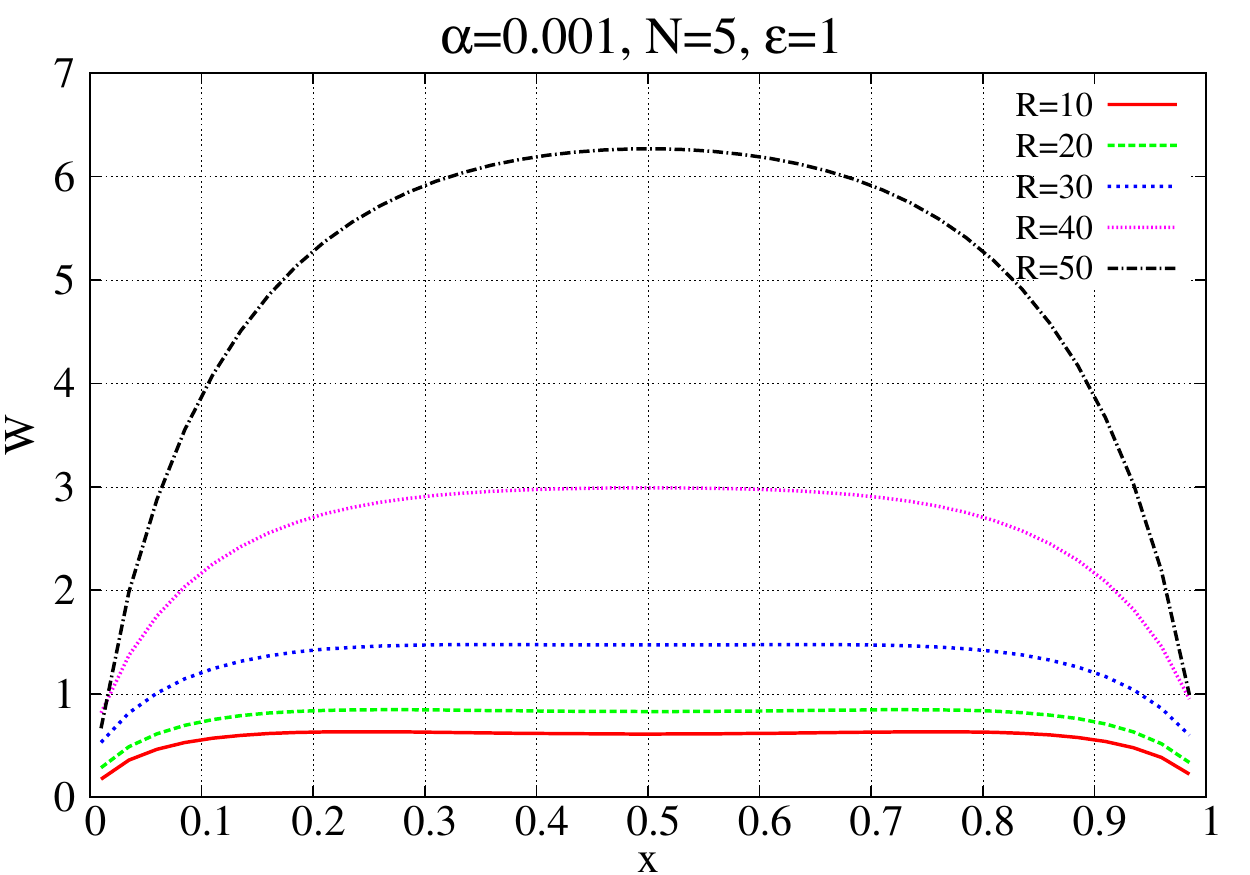}
  \caption{ Average work extraction as function of initial position of the partition for different $R$ values.}
  \label{W_vs_N_sw_R}
 \end{center}
\end{figure}

\begin{figure}[H]
 \begin{center}
  \includegraphics[height=2.5 in,width=2.5 in]{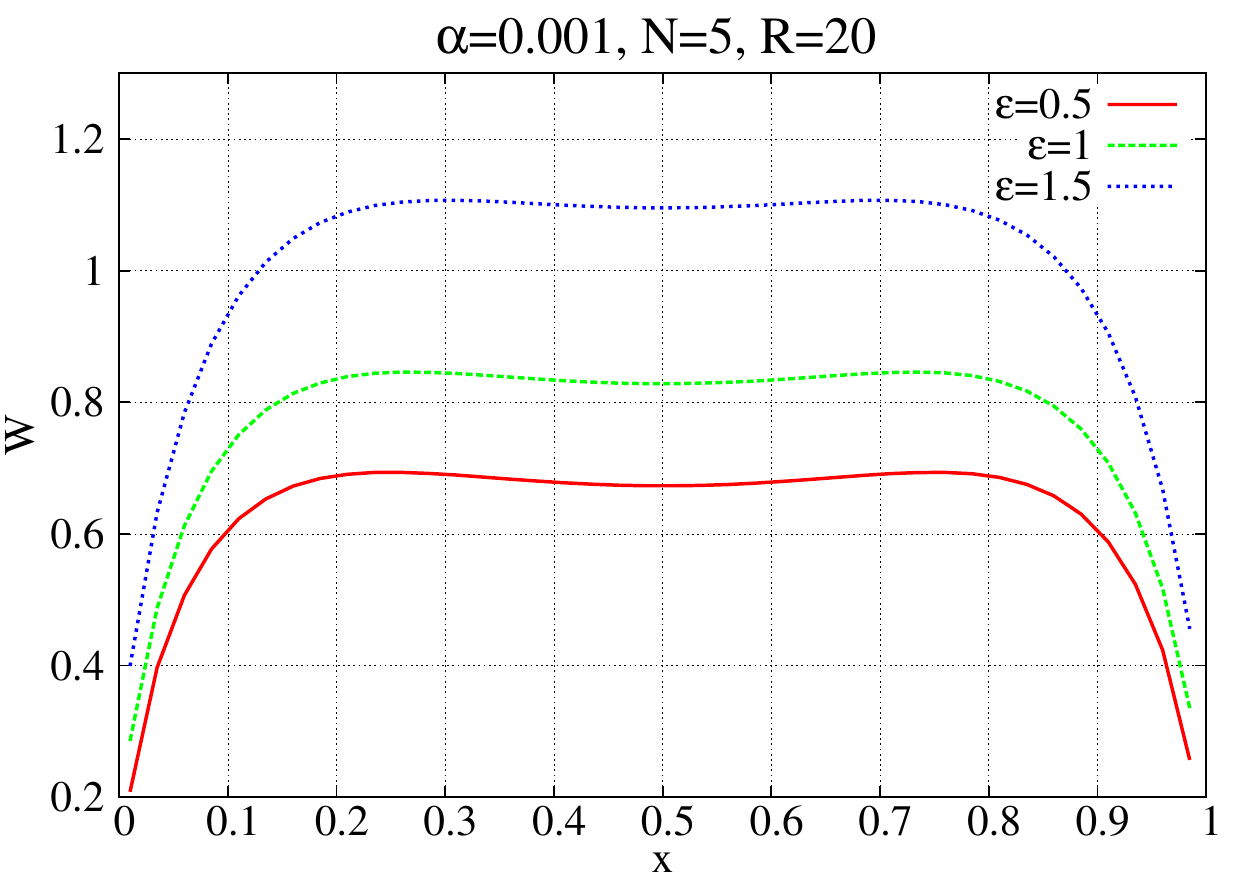}
  \caption{ Average work extraction as function of initial position of the partition for different values of well depth.}
  \label{W_vs_N_sw_ep}
 \end{center}
\end{figure}

Work extraction can be increased by varying the well width and depth.  Figs.\ref{W_vs_N_sw_R} and \ref{W_vs_N_sw_ep} show the behavior of work extraction when well width and depth is varied respectively.
More work is extracted as one increases the width and the depth of the well. 

\section{Discussions}
\label{dis}
\begin{figure}[H]
 \begin{center}
  \includegraphics[height=2.5 in,width=2.5 in]{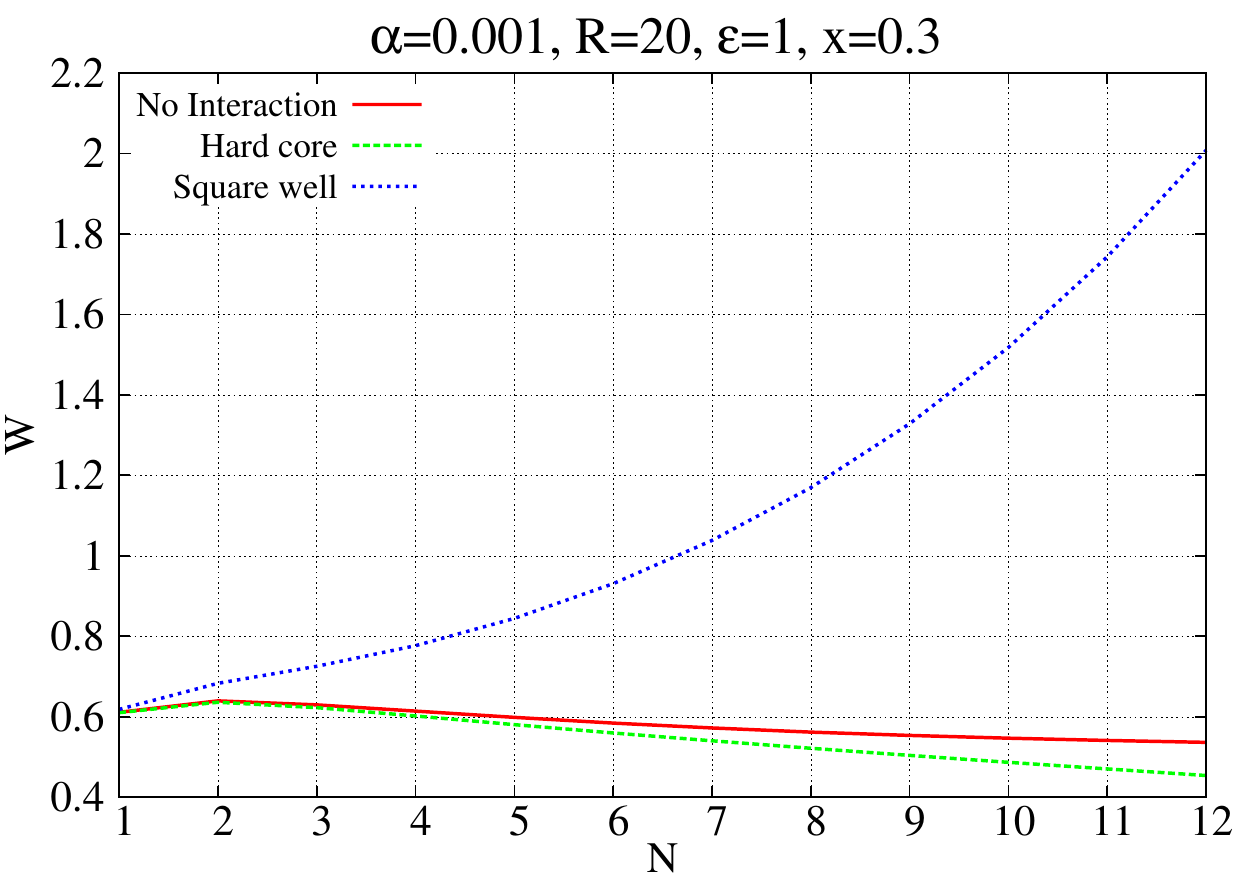}
  \caption{ Comparison of average work extraction as function of number of particles.}
  \label{compare1}
 \end{center}
\end{figure}

\begin{figure}[H]
 \begin{center}
  \includegraphics[height=2.5 in,width=2.5 in]{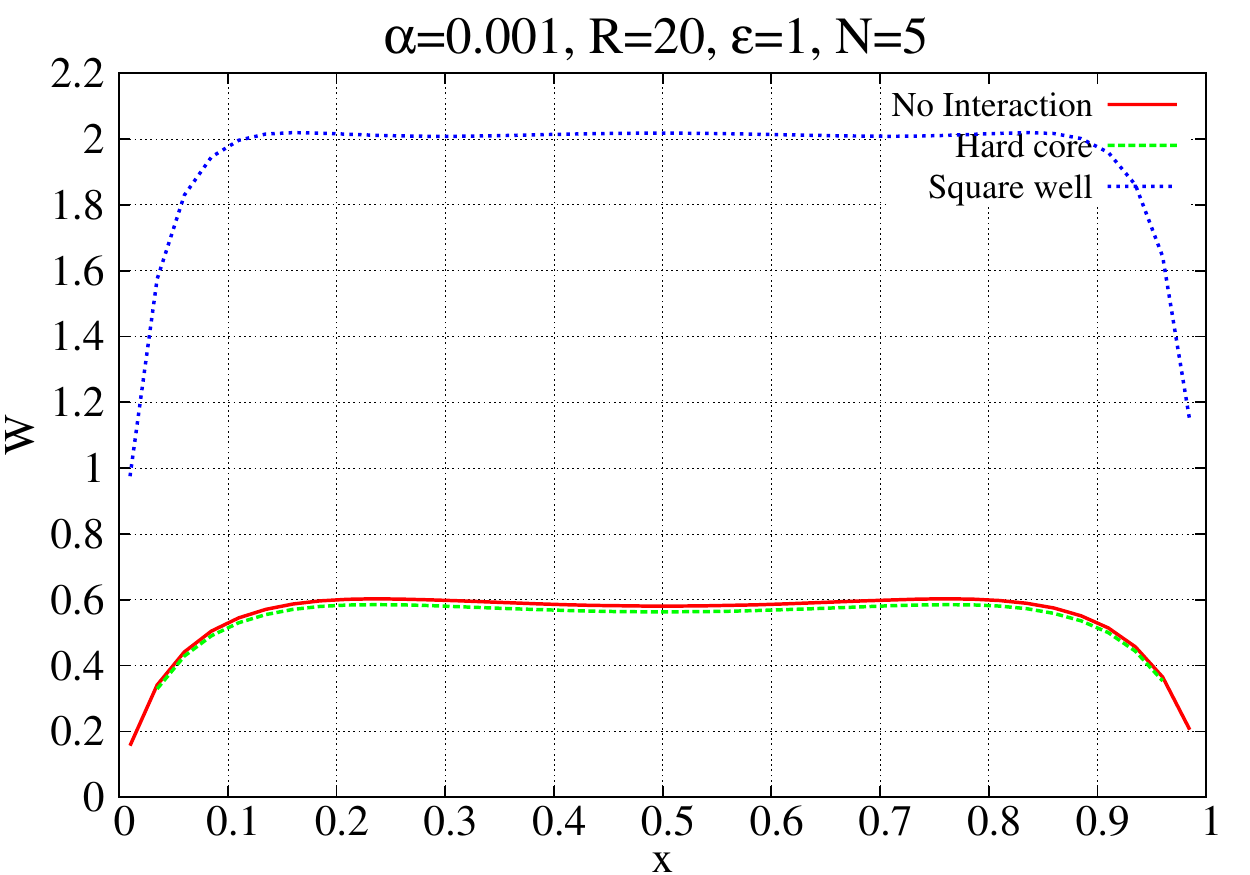}
  \caption{ Comparison of average work extraction as function of position of the partition.}
  \label{compare2}
 \end{center}
\end{figure}

We now discuss the role of interactions in work extraction by providing a comparative study between non-interacting and interacting particle systems. Figs. \ref{compare1} and \ref{compare2} compare the 
work extraction for three cases. Work extraction in systems with square well interactions between particles  is more with respect to non-interacting system and particles with  hard core interactions. 
This is due to the formation of bound states. If probability of formation of bound states is large due to increased well width and depth, there will be higher probability to find large number of particles
on one side. As a result the distance between the initial and the final equilibrium position of the partition will be more. This will in turn increase the work extraction. This effect is not seen in 
case of particles interacting via hard core interaction and hence work extraction decreases in that case.

\section{Conclusion}
\label{conclusion}
In conclusion,  we have studied the interacting multi-particle classical Szilard engine. In our article, we dealt with working systems consisting of particles which interact with themselves via hard core 
and square well interactions. A major assumption in our work is that we considered the working systems to be non-ideal gas with  low density, so that we can safely use the equation of state of the gas 
upto second virial co-efficient. One can go beyond this assumption to mimic with the realistic picture and consider higher order virial co-efficients but there is no end to it. It will obviously give
changes in work extraction. But to first approximation we are able to show analytically how inter-particle interactions affect work extraction. Our study reveals that work extraction is less when more
particles are interacting through hard core interactions. More work is extracted when the particles are interacting via square well interaction. Another important result for the second case(square well 
interactions)  is that as we increase the particle number the work extraction becomes independent of the initial position of the partition, as opposed to the first case(hard core interactions).
Work extraction depends crucially on the initial position of the partition. More work can be  extracted with larger number of particles when partition is inserted at positions near the boundary walls.

\section{Acknowledgement}
 AMJ thanks DST, India for financial support (through J. C. Bose National Fellowship).

\end{document}